\documentstyle[preprint,aps,epsfig]{revtex}

\begin{document}
\draft
\title{Anomalous Action \\in Gauge Invariant,
Nonlocal, Dynamical Quark Model}

\author{Yong-Liang Ma$^a$, Qing Wang$^{a,b}$}

\address{$^a$Department of Physics, Tsinghua University, Beijing 100084, China
\footnote{Mailing address}\\
$^b$Institute of Theoretical Physics, Academia Sinica, Beijing 100080, China}
%\date{Jan 24, 2003}

%
\maketitle
\begin{abstract}
Anomalous sector of chiral Lagrangian is calculated in
a gauge invariant, nonlocal, dynamical quark model. The Wess-Zumino term
is proved  coming from two kinds of sources, one is
 independent of and another dependent on dynamical quark self energy
 $\Sigma(k^2)$. $p^6$ and more higher order anomalous sectors come only from
 $\Sigma(k^2)$ dependent source. After some cancellation, standard
 Wess-Zumino action is obtained.
\end{abstract}

\bigskip
PACS number(s): 11.30.Rd, 12.38.Aw,12.38.Lg,12.39.Fe
\bigskip

%%%%%%%%%%%%%%%%%%%%%%%%

\vspace{1cm}
Chiral Lagrangian has successfully described low-energy hadronic process.
Within context of chiral Lagrangian approach that incorporates symmetries
of QCD, to a certain order in the low-energy expansion,
  the difference
 between different underlying theories with same spontaneously broken chiral
 symmetry is in the values of coefficients in the chiral Lagrangian.
 To test QCD in terms of chiral Lagrangian
 at quantitative level, we need to calculate values of these
 coefficients based on QCD and compare them with those from experiment data.
 Recently  chiral Lagrangian is exactly derived from underlying QCD and
 coefficients in
 the chiral Lagrangian are formally expressed in terms of Green's functions of
 underlying QCD \cite{WQ0}. Further it was shown that the
 coefficients for normal part with even intrinsic parity
 of pseudoscalar meson chiral Lagrangian is
  saturated by dynamical quark self energy $\Sigma(k^2)$\cite{GND}.
  While the anomaly contribution \cite{anomalyderive}, conventionally taken as
 the main source of the coefficients, are completely cancelled, leaving the
  $\Sigma(k^2)$ dependent coefficients which vanish when the strong
  interaction is switched off. Although this result is obtained in a special
 gauge invariant, nonlocal, dynamical (GND) quark model, considering this
 model can  be seen as a simple approximation for the exact derivation of
 chiral
  Lagrangian from underlying QCD \cite{WQ1}, the result is general.
  In this paper, we generalize the discussion for even intrinsic parity part of
   chiral Lagrangian in GND quark model to the anomalous part with odd
   intrinsic parity of chiral Lagrangian.

Anomalous part of chiral Lagrangian was first derived by Wess and
Zumino \cite{WZ} as a series expansion by directly integrating the
anomaly consistency conditions. This Lagrangian was later obtained
more directly by Witten \cite{Witten} by writing an action of
abnormal intrinsic parity with a free integer parameter fixed from
QCD anomaly. It was subsequently shown by several authors that
quantum corrections to the Wess-Zumino classical action do not
renormalize the coefficient of the $O(p^4)$ Wess-Zumino term and the
one-loop counter terms lead to conventional chiral invariant
 structures at $O(p^6)$ \cite{p60}. Ref.\cite{p61}
 classified all possible terms of chiral
 Lagrangian at order of $p^6$.

 The anomalous sector of chiral
 Lagrangian has already been discussed in the literature in quite some detail
 \cite{anomsum}, its derivation can be classified into two classes:
 one is related to integrating back infinitesimal variation of quark functional
  integration measure  \cite{WZ,R}.
  We call it the strong interaction dynamics independent approach,
  since except feature of mathematical beauty, this approach lacks  explicit
  correspondence with interaction.
    i.e. switching off strong interaction among quarks and gluons seems has no
    interference to the calculation procedure which only depends on
   transformation property of functional integration measure and couplings
   among  external fields and quarks. Therefore strong interaction dynamics in
   this approach has nothing to do with anomalous sector of chiral Lagrangian.
    This problem become serious since in GND quark
    model \cite{GND} and QCD \cite{WQ1}, such kind of terms independent of
    strong interaction are already proved to be exactly cancelled in normal
    part of chiral Lagrangian. Whether this
    cancellation continue to happen at anomalous sector of chiral Lagrangian?
 If so, we need to know whether there still has Wess-Zumino term?
if not, we need to specify the reason to break this cancellation.
Another type of derivation depends on
   strong interaction dynamics through constituent quark mass $M_Q$
   \cite{anomsum,quarkloop}.
 Anomalous sector in this approach
 which vanishes when we switch off strong interaction
 ($M_Q=0$) can be calculated either by directly performing loop calculations or
 computing $M_Q$ dependent
 determinants of quarks by choice of some proper regularization method.
 Since a hard constituent quark mass in the theory is only
 a rough approximation of QCD
 which causes wrong bad ultraviolet behavior of the theory, a more precise
 description should be replacing it with momentum dependent quark self energy
 $\Sigma(k^2)$ which plays an important role in QCD low energy physics \cite{PS}.
  In terms of $\Sigma(k^2)$, Ref.\cite{Cahill} expressed Wess-Zumino action in
  terms of quark self energy, but did not explain the relation of their result
  with that from dynamics independent approach. This relation must be
  clarified to avoid the double counting of anomalous sector from two
  different
  sources mentioned above. Since the relation for normal part from different
  approaches are already discussed in Ref.\cite{GND},
  in this work we focus on how
  to calculate anomalous  part of chiral Lagrangian in GND quark
  model as a QCD motivated discussion in which we will express Wess-Zumino
  action in terms of $\Sigma(k^2)$ and setup its relation with
  dynamics independent approach. We will show that there are two cancellation
  mechanisms, different choices of cancellation mechanisms will lead to
   different, dynamics independent or dependent, approaches of
 anomalous sector of chiral Lagrangian.

  We start from the GND quark model action in Euclidean space
  \footnote{We have included normalization factor
  $N'\equiv {\rm Tr}~ln[i\partial\!\!\!\! /\;+J]$ appeared in (13) but
  dropped in final expression (18) of Ref.\cite{GND} which is discussed in
  Minkovski space.}
 \cite{GND},
\begin{eqnarray}
S_{\rm GND}[U,J]\equiv
{\rm Tr}~ln[\partial\!\!\!\!/\;+J_{\Omega}+\Sigma(-\overline{\nabla}^2)]
-{\rm Tr}~ln[\partial\!\!\!\! /\;+J_{\Omega}]
+{\rm Tr}~ln[\partial\!\!\!\! /\;+J]
\label{SGND}\;.
\end{eqnarray}
Where Tr is trace for color, flavor, Lorentz spinor and space-time indices.
$J_{\Omega}$ is rotated external field which can be decomposed into
scalar, pseudoscalar, vector and axial vector parts, it relate to original
source $J$ through a local chiral rotation $\Omega(x)$
\begin{eqnarray}
J_{\Omega}(x)&\equiv& -iv\!\!\! /\;_{\Omega}(x)
-ia\!\!\! /\;_{\Omega}(x)\gamma_5-s_{\Omega}(x)+ip_{\Omega}(x)\gamma_5
\nonumber\\
&=&[\Omega(x)P_R+\Omega^{\dagger}(x)P_L]~[J(x)+\partial\!\!\!\! /\;_x]
~[\Omega(x)P_R+\Omega^{\dagger}(x)P_L]\;.
\label{JOmegadef}
\end{eqnarray}
The covariant differential $\overline{\nabla}^{\mu}$ is defined as
$\overline{\nabla}^{\mu}=\partial^\mu-iv_{\Omega}^\mu(x)$,
 local goldstone field $U(x)$ is related to rotation field $\Omega(x)$ by
$U(x)=\Omega^2(x)$.

The GND quark model given in (\ref{SGND}) is expressed on rotated basis, by
rotating back to unrotated basis, we get an alternative expression of it,
\begin{eqnarray}
S_{\rm GND}[U,J]= {\rm Tr}~ln[\partial\!\!\!\!/\;+J+\Pi]\label{SGNDo}\;,
\end{eqnarray}
with $\Omega$ and $\overline{\nabla}_{\mu}$ dependent $\Pi$ field
\begin{eqnarray}
\Pi(x)\equiv[\Omega^{\dagger}(x)P_R+\Omega(x)P_L]
\Sigma(-\overline{\nabla}^2)[\Omega^{\dagger}(x)P_R+\Omega(x)P_L]\;.
\end{eqnarray}

$S_{\rm GND}$ given in (\ref{SGNDo}) is explicitly dynamics dependent through
quark self energy $\Sigma(k^2)$, since goldstone field $\Omega$ couple to
external fields through $\Sigma(k^2)$, once $\Sigma(k^2)$ vanish, the
interaction disappear by $S_{\rm GND}\stackrel{\Sigma=0}{--\rightarrow}
{\rm Tr}~ln[\partial\!\!\!\!/\;+J]$, i.e., action becomes a pure external
field term.

To calculate the anomalous part of (\ref{SGNDo}), we introduce a parameter $t$
defined at interval $[0,1]$ and replace
original $\Omega(x)\equiv e^{i\pi(x)}$ appeared in $\Pi(x)$ of
(\ref{SGNDo}) by $\Omega(t,x)=e^{it\pi(x)}$ with
$\Omega(1,x)=\Omega(x)$ and $\Omega(0,x)=1$. Correspondingly,
$\Pi(x)$ is replaced by $\Pi(t,x)$ with $\Pi(1,x)=\Pi(x)$ and
$\Pi(0,x)= \Sigma[-(\partial_{\mu}-iv_{\mu}(x))^2]$. This together
with $\delta{\rm lnDet}A={\rm Tr}\delta AA^{-1}$ leads to
\begin{eqnarray}
&&S_{\rm GND}[U,J]-S_{\rm GND}[1,J]\nonumber\\
&&=\int_0^1 dt~{\rm Tr}~\bigg[\frac{\partial\Pi(t)}{\partial t}
[\partial\!\!\!\!/\;+J+\Pi(t)]^{-1}\bigg]\nonumber\\
&&=N_c\int_0^1 dt\int d^4x\int\frac{d^4k}{(2\pi)^4}~{\rm tr}_{lf}
~\bigg[\frac{\partial\Pi(t,k,x)}{\partial t}
[\partial\!\!\!\!/\;_x-ik\!\!\! /\;+J(x)+\Pi(t,k,x)]^{-1}\bigg]\;,\label{Cal0}
\end{eqnarray}
where $S_{\rm GND}[1,J]$ can be dropped out , since it is a $U$
field independent pure external field term. tr$_{lf}$ is trace for
Lorentz spinor and flavor indices and
\begin{eqnarray}
\Pi(t,k,x)\equiv\bigg[[\Omega^{\dagger}(x)P_R+\Omega(x)P_L]
\Sigma[-(\overline{\nabla}_x-ik)^2][\Omega^{\dagger}(x)P_R+\Omega(x)P_L]
\bigg]_{\Omega(x)\rightarrow\Omega(t,x)}\;.\label{Pikdef}
\end{eqnarray}
(\ref{Cal0}) includes all interaction terms in the GND quark
model. We are only interested in its anomalous part which can be
shown that it is the part containing only one Levi-Civita tensor
\cite{anomsum}, we use subscript $\epsilon$ to specify this
operation of keeping one $\epsilon_{\mu\nu\mu'\nu'}$ tensor, then
\begin{eqnarray}
&&S_{\rm GND}[U,J]\bigg|_{\rm anomalous~part}\nonumber\\
&&=N_c\int_0^1 dt\int d^4x\int\frac{d^4k}{(2\pi)^4}~{\rm tr}_{lf}
~\bigg[\frac{\partial\Pi(t,k,x)}{\partial t}
[\partial\!\!\!\!/\;_x-ik\!\!\! /\;+J(x)+\Pi(t,k,x)]^{-1}\bigg]_{\epsilon}\;.
\label{Cal1}
\end{eqnarray}
The next step we are interested in
is to calculate the leading order contribution
without external fields. We switch off the external field by taking $J=0$, then
$\overline{\nabla}_x^{\mu}$ in (\ref{Pikdef}) becomes
$\overline{\nabla}_x^{\mu}=\partial_x^{\mu}
+\frac{1}{2}[\Omega^{\dagger}(t,x)\partial^{\mu}\Omega(t,x)+
\Omega(t,x)\partial^{\mu}\Omega^{\dagger}(t,x)]$. We can take momentum
(or derivatives) expansion, scale external field $s$ and $p$ as order
of $p^2$, $v_{\mu}$ and $a_{\mu}$ as order of $p^1$, $U$ and $\Omega$ as order
of $p^0$. Then, up to the order of $p^4$, after
detail calculation, we find
\begin{eqnarray}
&&S_{\rm GND}[U,J]\bigg|_{{\rm anomalous~part},~J=0,~p^4}\equiv\Gamma^-[U]
\nonumber\\
&&=-\frac{N_c C}{48\pi^2}\int_0^1 dt\int d^4x~
\epsilon^{\mu\nu\mu'\nu'}{\rm tr}_f\bigg[U^{\dagger}(t,x)
\frac{\partial U(t,x)}{\partial t}L_{\mu}(t,x)L_{\nu}(t,x)
L_{\mu'}(t,x)L_{\nu'}(t,x)\bigg]\;,\label{WZ0}
\end{eqnarray}
with $L_{\mu}(t,x)\equiv U^{\dagger}(t,x)\partial_{\mu}U(t,x)$ and
\begin{eqnarray}
C&=&\frac{3}{\pi^2}\int d^4k\bigg[
\frac{4\Sigma^6(k^2)}{[k^2+\Sigma^2(k^2)]^5}
-\frac{8k^2\Sigma^{\prime}(k^2)\Sigma^5(k^2)}{[k^2+\Sigma^2(k^2)]^5}
\bigg]\nonumber\\
&=&-12\int_0^{\infty} dk^2
\bigg[\frac{k^2\Sigma^4(k^2)}{[k^2+\Sigma^2(k^2)]^3}\frac{d}{dk^2}
\frac{\Sigma^2(k^2)}{[k^2+\Sigma^2(k^2)]}\bigg]=1\label{Ccal}
\end{eqnarray}
where \footnote{Result given in Ref.\cite{Cahill} is  different with us in the
integrand by an extra factor $k^2/[k^2+\Sigma^2(k^2)]$.}
 by changing the integration variable to $t=\Sigma^2(k^2)/k^2$, the
integration can be finished
 with  result value 1 in the case of infinity upper limit of momentum
 integration and requiring  quark self energy satisfies  following constraint
 \begin{eqnarray}
 \frac{\Sigma(k^2)}{k^2}\rightarrow\left\{\begin{array}{lcl}
 \infty &~~~&k^2=0 \\ 0 &&k^2\rightarrow\infty\end{array}\right.
 \label{constraint}
 \end{eqnarray}
   Note constant $C$ although is $\Sigma(k^2)$ dependent, but in a rather wide
  range of nonzero $\Sigma(k^2)$ given by (\ref{constraint}) is 1,
 \begin{eqnarray}
 C=\left\{\begin{array}{lcl}
 1 &~~~&\Sigma(k^2)\neq 0 \\ 0 &&\Sigma(k^2)=0\end{array}\right.
 \end{eqnarray}

 In a five dimension disc Q with its coordinates
$y^i (i=1\ldots 5)$ and  4-dimension space-time boundary, we have
\begin{eqnarray}
&&\frac{\partial}{\partial t}{\rm tr}_f\bigg[
L_i(t,y)L_j(t,y)L_k(t,y)L_l(t,y)L_m(t,y)\bigg]\epsilon^{ijklm}\nonumber\\
&&=5\frac{\partial}{\partial y^m}{\rm tr}_f\bigg[
U^{\dagger}(t,y)\frac{\partial U(t,y)}{\partial t}L_i(t,y)L_j(t,y)
L_k(t,y)L_l(t,y)\bigg]\epsilon^{ijklm}\;,\nonumber
\end{eqnarray}
then
\begin{eqnarray}
&&\int_Q d\Sigma^{ijklm}{\rm tr}_f[L_i(1,y)L_j(1,y)L_k(1,y)L_l(1,y)
L_m(1,y)]\nonumber\\
&&=\int d\Sigma^{ijklm}\int_0^1dt~\frac{\partial}{\partial t}{\rm tr}_f[
L_i(t,y)L_j(t,y)L_k(t,y)L_l(t,y)L_m(t,y)]\nonumber\\
&&=5\int d^4x\int_0^1dt~\epsilon^{\mu\nu\sigma\rho} {\rm
tr}_f\bigg[ U^{\dagger}(t,x)\frac{\partial U(t,x)}{\partial
t}L_{\mu}(t,x)L_{\nu}(t,x) L_{\sigma}(t,x)L_{\rho}(t,x)\bigg]\;.\nonumber
\end{eqnarray}
This relation leads (\ref{WZ0}) to Wess-Zumino action
\begin{eqnarray}
\Gamma^-[U]&=&-\frac{N_c}{240\pi^2}\int_Q
 d\Sigma^{ijklm}{\rm tr}_f[L_i(1,y)L_j(1,y)L_k(1,y)L_l(1,y)
L_m(1,y)]\;.\label{WZf}
\end{eqnarray}

In principle, we can follow the present procedure given by
(\ref{Cal1}) to continue the calculation for external fields
dependent $p^4$ order terms, $p^6$ order terms, $\ldots$, etc. We
will discuss external fields dependent $p^4$ order terms in
follows in an alternative way and leave the detail discussion of those
more higher order lengthy results else where.

Now we discuss the chiral transformation property of $S_{\rm GND}$. Consider
following  $U_L(3)\otimes U_R(3)$ local chiral transformation with left and
right hands transformation matrices $V_L(x)$ and $V_R(x)$
\begin{eqnarray}
J(x)&\rightarrow& J'(x)=[V_R(x)P_L+V_L(x)P_R][J(x)+\partial\!\!\!\! /\;_x]
[V_R^{\dagger}(x)P_R+V_L^{\dagger}(x)P_L]\nonumber\\
\Omega(x)&\rightarrow&\Omega'(x)=h(x)\Omega(x)V_L^{\dagger}(x)
=V_R(x)\Omega(x)h^{\dagger}(x)\;,\label{Omegatrans}
\end{eqnarray}
with $h(x)$ depends on $V_R$, $V_L$ and $\Omega$, represents an induced hidden
local $U(3)$ symmetry to keep transformed $\Omega$ be a representative element
at coset class. In Ref. \cite{GND}, it was shown that the corresponding
transformation property for $J_{\Omega}$ and $\Sigma(-\overline{\nabla}^2)$ are
\begin{eqnarray}
&&J_{\Omega}(x)\rightarrow J_{\Omega}'(x)
=h(x)[J_{\Omega}(x)+\partial\!\!\!\! /\;_x]h^{\dagger}(x)\nonumber\\
&&\Sigma(-\overline{\nabla}^2)\rightarrow
\Sigma(-{\overline{\nabla}'}^2)
=h(x)\Sigma(-\overline{\nabla}^2)h^{\dagger}(x)\;.\label{Sigmatrans}
\end{eqnarray}
With (\ref{Omegatrans}) and (\ref{Sigmatrans}), we find $\Pi$ field transforms
as
\begin{eqnarray}
\Pi(x)\rightarrow \Pi'(x)=[V_R(x)P_L+V_L(x)P_R]
\Pi(x)[V_R^{\dagger}(x)P_R+V_L^{\dagger}(x)P_L]\;.
\end{eqnarray}
Then for infinitesimal transformation,
\begin{eqnarray}
V_R(x)=1+i\alpha(x)+i\beta(x)+\cdots\hspace{2cm}
V_L(x)=1+i\alpha(x)-i\beta(x)+\cdots\;\;,\nonumber
\end{eqnarray}
we have
\begin{eqnarray}
&&\delta[\partial\!\!\!\!/\;+J]=
i[\alpha(x)-\beta(x)\gamma_5][\partial\!\!\!\!/\;+J]
-i[\partial\!\!\!\!/\;+J][\alpha(x)+\beta(x)\gamma_5]\nonumber\\
&&\delta\Pi(x)=i[\alpha(x)-\beta(x)\gamma_5]\Pi(x)
-i\Pi(x)[\alpha(x)+\beta(x)\gamma_5]\;,\nonumber
\end{eqnarray}
the corresponding infinitesimal transformation of $S_{\rm GND}$ is
\begin{eqnarray}
\delta S_{\rm GND}
&=&{\rm Tr}\bigg[
[\partial\!\!\!\!/\;+J+\Pi]^{-1}\delta[\partial\!\!\!\!/\;+J+\Pi]
\bigg]\nonumber\\
&=&i{\rm Tr}\bigg[
[\partial\!\!\!\!/\;+J+\Pi]^{-1}\bigg(
[\alpha(x)-\beta(x)\gamma_5][\partial\!\!\!\!/\;+J+\Pi]
-[\partial\!\!\!\!/\;+J+\Pi][\alpha(x)+\beta(x)\gamma_5]\bigg)
\bigg]\nonumber\\
&=&-2i{\rm Tr}[\beta\gamma_5]\nonumber\\
&=&-2i\lim_{\Lambda\rightarrow\infty}{\rm Tr}\bigg[\beta\gamma_5
e^{\frac{[\partial\!\!\!\!/\;+J+\Pi]^2}{\Lambda^2}}\bigg]\;,\label{variS0}
\end{eqnarray}
where we have taken the Fujikawa approach \cite{Fuji} in last
equality to regularize the anomaly. The detail calculation shows
that the $\Pi$ field, scalar and pseudoscalar part of external
fields on the exponential of regulator make no contribution to the
result in the limit of infinite $\Lambda$. The result of
calculation for (\ref{variS0}) just gives the standard Bardeen anomaly
\cite{Bardeen},
\begin{eqnarray}
\delta S_{\rm GND}&=&-i\int d^4x~{\rm tr}_f[\beta(x)\tilde{\Omega}(x)]
\label{variS}\\
\tilde{\Omega}(x)&=&{N_c\over 16\pi^2}~\epsilon^{\mu\nu\mu'\nu'}~\bigg\{
V_{\mu\nu}(x)V_{\mu'\nu'}(x)+{4\over3}d_\mu
a_\nu(x) d_{\mu'} a_{\nu'}(x)\label{tOmegadef}\\
&&+{2i\over3}\{V_{\mu\nu}(x),a_{\mu'}(x)a_{\nu'}(x)\}
+{8i\over3}a_\mu(x) V_{\mu'\nu'}(x)a_\nu(x)+{4\over3}a_\mu(x)
a_\nu(x) a_{\mu'}(x)a_{\nu'}(x)\bigg\}\;,\nonumber
\end{eqnarray}
where
$V_{\mu\nu}=\partial_{\mu}v_{\nu}-\partial_{\nu}v_{\mu}-i[v_{\mu},v_{\nu}]$
and $d_{\mu}a_{\nu}=\partial_{\mu}a_{\nu}-i[v_{\mu},a_{\nu}]$

With result (\ref{variS}), we can follow the procedure given by Wess and Zumino  \cite{WZ} to construct corresponding $p^4$ order
anomalous part action by taking a special local
chiral rotation $\delta_{\beta}=\delta\bigg|_{\alpha=0}$, and
\begin{eqnarray}
U(x)=e^{-2i\beta(x)}\hspace{2cm}  e^{\delta_{\beta}} U(x)=1\;,
\end{eqnarray}
then rotation of action (\ref{SGNDo}) leads
\begin{eqnarray}
e^{\delta_{\beta}}S_{\rm GND}[U,J]=
{\rm Tr}~ln[\partial\!\!\!\!/\;+J_{\Omega}+\Sigma(-\overline{\nabla}^2)]
\bigg|_{\Omega(x)=e^{-i\beta(x)}}\;,
\end{eqnarray}
therefore
\begin{eqnarray}
S_{\rm GND}[U,J]&=&
\bigg[\frac{1-e^{\delta_{\beta}}}{\delta_{\beta}}\delta_{\beta}S_{\rm GND}[U,J]
+{\rm Tr}~ln[\partial\!\!\!\!/\;+J_{\Omega}+\Sigma(-\overline{\nabla}^2)]
\bigg]_{\Omega(x)=e^{-i\beta(x)}}\nonumber\\
&=&\bigg[i\int_0^1dt\int d^4x~e^{t\delta_{\beta}}
~{\rm tr}_f[\beta(x)\tilde{\Omega}(x)]
+{\rm Tr}~ln[\partial\!\!\!\!/\;+J_{\Omega}+\Sigma(-\overline{\nabla}^2)]
\bigg]_{\Omega(x)=e^{-i\beta(x)}}\;\label{SGNDWZ}
\end{eqnarray}
where the first term is the standard Wess-Zumino term constructed by Wess and
Zumino in \cite{WZ}. While for the second term, vector-like transformation law
(\ref{Sigmatrans}) imply it is invariant under chiral rotation
(\ref{Omegatrans}) and then there will be no Wess-Zumino term in it, since
Wess-Zumino term is not invariant under chiral rotation, it gives chiral
anomaly \cite{WZ}. We can directly verify this by performing  the similar
computation procedure given in (\ref{Cal1}),
the contribution to $p^4$ order anomalous term  from $\Sigma$ dependent term of
 Tr$ln[\partial\!\!\!\!/\;+J_{\Omega}+\Sigma(-\overline{\nabla}^2)]$ is
 \begin{eqnarray}
&&{\rm Tr}ln[\partial\!\!\!\!/\;+J_{\Omega}+\Sigma(-\overline{\nabla}^2)]
\bigg|_{\Sigma~{\rm dependent, anomalous}~p^4}\nonumber\\
&&=-2N_c\epsilon_{\mu\nu\alpha\beta}\int_0^1
dt\int\frac{d^4k}{(2\pi)^4}{\rm tr_f}
\bigg\{{\partial U\over\partial
t}U^\dag\bigg[\{\frac{\Sigma^2(k^2)[\Sigma^2(k^2)-k^2]}{[\Sigma^2(k^2)+k^2]^4}
-\frac{2k^2\Sigma^\prime(k^2)\Sigma(k^2)[\Sigma^2(k^2)-k^2]}
{[\Sigma^2(k^2)+k^2]^4}\}\nonumber\\
&&\hspace{0.5cm}\times(
2\bar{\nabla}_\mu\bar{\nabla}_\nu\bar{\nabla}_\alpha\bar{\nabla}_\beta
+2a_\mu a_\nu\bar{\nabla}_\alpha\bar{\nabla}_\beta
-2\bar{\nabla}_\mu a_\nu\bar{\nabla}_\alpha a_\beta
+2\bar{\nabla}_\mu a_\nu a_\alpha\bar{\nabla}_\beta
+2a_\mu\bar{\nabla}_\nu \bar{\nabla}_\alpha a_\beta
-2a_\mu\bar{\nabla}_\nu a_\alpha\bar{\nabla}_\beta\nonumber\\
&&\hspace{0.5cm}+2\bar{\nabla}_\mu\bar{\nabla}_\nu a_\alpha a_\beta
+2a_\mu a_\nu a_\alpha a_\beta)
+\{\frac{k^2\Sigma^2(k^2)}{[\Sigma^2(k^2)+k^2]^4}
-\frac{2k^4\Sigma^\prime(k^2)\Sigma(k^2)}{[\Sigma^2(k^2)+k^2]^4}\}
(4\bar{\nabla}_\mu\bar{\nabla}_\nu\bar{\nabla}_\alpha\bar{\nabla}_\beta
\nonumber\\
&&\hspace{0.5cm}+2a_\mu a_\nu\bar{\nabla}_\alpha\bar{\nabla}_\beta
-2\bar{\nabla}_\mu a_\nu\bar{\nabla}_\alpha a_\beta
+4a_\mu\bar{\nabla}_\nu \bar{\nabla}_\alpha a_\beta
-2a_\mu\bar{\nabla}_\nu a_\alpha\bar{\nabla}_\beta
+2\bar{\nabla}_\mu\bar{\nabla}_\nu a_\alpha a_\beta)\bigg]\bigg\}\nonumber\\
&&=\frac{1}{2}
\int_0^1 dt\int d^4x~\bigg[\frac{\partial U}{\partial t}U^{\dagger}
\tilde{\Omega}(t,x)\bigg]\;,\label{cancel}
\end{eqnarray}
where as done in (\ref{Ccal}), the momentum integration in above
formulae for $\Sigma$ dependent coefficients can be finished,
result in a Bardeen anomaly expressed in terms of rotated external
fields in the case of infinity upper limit of momentum
integration.
 $\tilde{\Omega}(t,x)$ is
  $\tilde{\Omega}(x)$ defined in (\ref{tOmegadef}) with
 all $\Omega(x)$ replaced by $\Omega(t,x)$.
  This term exactly cancels $\Sigma$ independent term of
 Tr$ln[\partial\!\!\!\!/\;+J_{\Omega}+\Sigma(-\overline{\nabla}^2)]$
  which is  Tr$ln[\partial\!\!\!\!/\;+J_{\Omega}]$ and
    we will calculate minus of it later in (\ref{Sp4WZ1}). So
our result shows that there is no $p^4$ order anomalous term in
 Tr$ln[\partial\!\!\!\!/\;+J_{\Omega}+\Sigma(-\overline{\nabla}^2)]$,
 the contribution to anomalous part from this term is at least
order of $O(p^6)$ \footnote{Since this term in original Wess-Zumino paper
\cite{WZ} is forced to be unity as a normalization constant, the $O(p^6)$
anomalous term is dropped out there.}.

 Combine two terms in (\ref{SGNDWZ}) together,
$p^4$ and more higher order anomalous terms are all
 included in $S_{\rm GND}[U,J]$.

Compare result (\ref{SGNDWZ}) with (\ref{SGND}), we find the last two terms in
 (\ref{SGND}) is
\begin{eqnarray}
-{\rm Tr}~ln[\partial\!\!\!\! /\;+J_{\Omega}]
+{\rm Tr}~ln[\partial\!\!\!\! /\;+J]
=i\int_0^1dt\int d^4x~e^{-t\delta_{\beta}}
~{\rm tr}_f[\beta(x)\tilde{\Omega}(x)]
\bigg|_{\Omega(x)=e^{-i\beta(x)}}\;.\label{Sp4}
 \end{eqnarray}
This shows that the last two terms in (\ref{SGND}) only contribute $p^4$ order
 anomalous terms which are Wess-Zumino action.
 In fact, we can  explicitly get Wess-Zumino action from l.h.s of (\ref{Sp4})
 by taking similar computation procedure in (\ref{Cal1}).
 Replacing  $\Omega(x)$  by $\Omega(t,x)=e^{it\pi(x)}$, we can show that
under infinitesimal variation of parameter $t\rightarrow t+\delta t$,
$\delta [\partial\!\!\!\! /\;+J_{\Omega}(t)]=i\delta t\{\pi(x)\gamma_5,
 [\partial\!\!\!\! /\;+J_{\Omega}(t)]\}$, which leads,
\begin{eqnarray}
-{\rm Tr}~ln[\partial\!\!\!\! /\;+J_{\Omega}]
+{\rm Tr}~ln[\partial\!\!\!\! /\;+J]
&=&-\int_0^1dt~{\rm Tr}\bigg[\frac{\partial J_{\Omega(t)}}{\partial t}
[\partial\!\!\!\! /\;+J_{\Omega(t)}]^{-1}\bigg]\nonumber\\
&=&-\lim_{\Lambda\rightarrow\infty}
\int_0^1 dt~{\rm Tr}\bigg[\frac{\partial U}{\partial t}U^{\dagger}
\gamma_5e^{\frac{[\partial\!\!\!\! /\;+J_{\Omega(t)}]^2}{\Lambda^2}}\bigg]\;.
\end{eqnarray}
Since we are interested in anomalous part, we only need to collect the
one $\epsilon_{\mu\nu\mu'\nu'}$ dependent terms of above result, the detail
calculation gives
\begin{eqnarray}
&&\bigg[-{\rm Tr}~ln[\partial\!\!\!\! /\;+J_{\Omega}]
+{\rm Tr}~ln[\partial\!\!\!\! /\;+J]\bigg]_{\rm anomalous~part}
\nonumber\\
&&=-\lim_{\Lambda\rightarrow\infty}
\int_0^1 dt~{\rm Tr}\bigg[\frac{\partial U}{\partial t}U^{\dagger}
\gamma_5e^{\frac{[\partial\!\!\!\! /\;+J_{\Omega(t)}]^2}{\Lambda^2}}
\bigg]_{\epsilon}
=\frac{1}{2}
\int_0^1 dt\int d^4x~\bigg[\frac{\partial U}{\partial t}U^{\dagger}
\tilde{\Omega}(t,x)\bigg]\;,\label{Sp4WZ1}
\end{eqnarray}
 As in (\ref{cancel}) $\tilde{\Omega}(t,x)$ is
  $\tilde{\Omega}(x)$ defined in (\ref{tOmegadef})
 by replacing all $\Omega(x)$ with $\Omega(t,x)$.
We further focus on zero external
fields terms. In this case, we can show that the $\Omega(t)$
dependent external fields satisfy constraints:
$d^{\mu}a_{\Omega}^{\nu}=d^{\nu}a_{\Omega}^{\mu}$ and
$V_{\Omega}^{\mu\nu}=i[a_{\Omega}^{\mu},a_{\Omega}^{\nu}]$. With
help of these relations, (\ref{Sp4WZ1}) can be further reduced to
$\Gamma^-[U]$ given in (\ref{WZ0}) with $C=1$ and with help of discussion
 after (\ref{WZ0}) we find it is just Wess-Zumino action with
 vanishing external fields.

 Up to now,  we have known
  that in GND quark model given by (\ref{SGND}), for anomalous sector,
 the last two terms contribute to $p^4$ order anomalous term, i.e. Wess-Zumino
  action, while the first term in  (\ref{SGND}) which is also the second term
  in  (\ref{SGNDWZ}), as discussed  previously, is invariant under chiral
  rotation and only contribute to $p^6$ or more higher  order anomalous terms.

The cancellation discussed for normal part in Ref.\cite{GND}
also happens here.
There are two kinds of cancellation mechanism now.
In the first cancellation
mechanism,
Wess-Zumino term contributed from $\Sigma$ independent and dependent parts of
first term of (\ref{SGND}) cancelled each  other, leaving only $O(p^6)$
anomalous terms. With this cancellation mechanism, the left
the Wess-Zumino term in GND quark model is from the second term of
(\ref{SGND}).
This is just the dynamics independent approach, since the source of result
Wess-Zumino action  here is independent of strong interaction dynamics
induced quark self  energy $\Sigma(k^2)$. In the second cancellation mechanism,
Wess-Zumino term given by second term of (\ref{SGND}) is completely cancelled
by the $\Sigma$ independent part of first term, leaving GND quark model with
the $\Sigma$ dependent part of first term as
Wess-Zumino action which we have explicitly expressed in (\ref{SGNDo}) and
calculated through formulae (\ref{Cal0}) to (\ref{WZf}). This is the
dynamics dependent approach. Both approaches generate same Wess-Zumino
term. Different approaches are corresponding to different
choices of cancellation mechanisms.

%%%%%%%%%%%%%%%%%%%%%%%%%%%%%%%%
\section*{Acknowledgments}

This work was  supported by National  Science Foundation of China No.90103008
and fundamental research grant of Tsinghua University.

%%%%%%%%%%%%%%%%%%%%%%%%%%%%%%%%%%%%%%%%%%%%%%%%%%%%%%%%%%%%%%%%%%%%%%%%%%%%%

\end{document}